\pdfoutput=1
\documentclass[a4paper,american,floatfix,pdftex,superscriptaddress,twoside,%
aps,prl,%
reprint,twocolumn,
]{revtex4-1}%
\usepackage{amsfonts,amsmath,amssymb}
\usepackage[T1]{fontenc}
\usepackage{graphicx}%
\usepackage[utf8]{inputenc}
\usepackage{microtype}
\usepackage[obeyFinal,textsize=footnotesize]{todonotes}
\usepackage{xspace}
\usepackage{hyperref, hypernat}
\usepackage[todos]{CMImacros}

\graphicspath{{./fig/}} 
\hypersetup{citebordercolor=yellow,linkbordercolor=red,urlbordercolor=blue}

\makeatletter%
\renewcommand*{\@fnsymbol}[1]{\ensuremath{\ifcase#1\or \|\or \mathsection\or \ddagger\or *\or **\or
      \mathparagraph\or \dagger\or \dagger\dagger \or \ddagger\ddagger \else\@ctrerr\fi}}%
\makeatother

\newcommand*{\cfeldesy}{\affiliation{Center for Free-Electron Laser Science, Deutsches
      Elektronen-Synchrotron DESY, Notkestraße 85, 22607 Hamburg, Germany}}%
\newcommand*{\uhhcui}{\affiliation{Center for Ultrafast Imaging, Universität Hamburg, Luruper
      Chaussee 149, 22761 Hamburg, Germany}}%
\newcommand*{\uhhphys}{\affiliation{Department of Physics, Universität Hamburg, Luruper Chaussee
      149, 22761 Hamburg, Germany}}%
\newcommand*{\jilinu}{\altaffiliation[Permanent address: ]{Institute of Atomic and Molecular
      Physics, Jilin University, Changchun 130012, China}}
\newcommand*{\tsinghua}{\altaffiliation[Permanent address: ]{Department of Physics, Tsinghua
      University, 100084, Beijing, China}}%
\newcommand*{\ucl}{\altaffiliation[Present address: ]{Department of Physics and Astronomy,
      University College London, Gower Street, London WC1E 6BT, UK}}%
\newcommand*{\jkemail}{\email[Corresponding author. Email:~]{jochen.kuepper@cfel.de}}%
\newcommand*{\cmiweb}{\homepage[URL:~]{https://www.controlled-molecule-imaging.org}}%

\begin{document}
\title{Spatial separation of 2-propanol monomer and its ionization-fragmentation pathways}%
\author{Jia Wang}\tsinghua\cfeldesy%
\author{Lanhai He}\jilinu\cfeldesy%
\author{Jovana Petrovic}\cfeldesy%
\author{Ahmed Al-Refaie}\ucl\cfeldesy%
\author{Helen~Bieker}\cfeldesy\uhhphys%
\author{Jolijn Onvlee}\cfeldesy\uhhcui%
\author{Karol Długołęcki}\cfeldesy%
\author{Jochen Küpper}\jkemail\cmiweb\cfeldesy\uhhphys\uhhcui%
\begin{abstract}\noindent%
   The spatial separation of 2-propanol monomer from its clusters in a molecular beam by an
   electrostatic deflector was demonstrated. Samples of 2-propanol monomer with a purity of 90~\%
   and a beam density of $7\times10^6~\text{cm}^{-3}$ were obtained. These samples were utilized to
   study the femtosecond-laser-induced strong-field multi-photon ionization and fragmentation of
   2-propanol using non-resonant 800~nm light with peak intensities of 3--$7\times10^{13}~\Wpcmcm$.
   \begin{center}
      \vspace{1ex}%
      \emph{In memoriam Jon T.\ Hougen}
   \end{center}
\end{abstract}
\maketitle

\section{Introduction}
2-propanol (C$_3$H$_8$O, also
isopropyl alcohol, isopropanol) is the simplest secondary alcohol. It possesses a structure with
three nonrigid internal rotations, \ie, of the hydroxyl group OH and two methyl tops
CH$_{3}$~\cite{Burenin:OS120:848, Burenin:OS120:854, Maeda:AJSS166:650}. 2-propanol has attracted
much attention, not only because it is highly valued as a preservative and used as an antiseptic in
the clinical environment, but also since it is widely used as an industrial solvent and cleaning
fluid, such as gasoline additive, an alkylating agent, and a disinfectant~\cite{Gavin:CD65:101,
   Dobrowolski:VS48:82}. The vibrational spectrum of 2-propanol has been studied in the early 1960s,
which suggested that 2-propanol in the gas phase exists in trans and gauche
conformations~\cite{Tanaka:NKZ83:521}. These stable isomers were confirmed by the
microwave~\cite{Kondo:JMS34:97, Hirota:JPC83:1457, Ulenikov:JMS145:262, Hirota:JMS207:243} and the
millimeter- and submillimeter-wave spectra~\cite{Maeda:AJSS166:650, Dobrowolski:VS48:82}.

Supersonic molecular beam is a valuable tool for molecular spectroscopy and the studies of
molecular dynamics and reactions~\cite{Scoles:MolBeam:1and2, Hillenkamp:JCP118:8699,
   Wang:RSC47:6744}. For instance, this was exploited in studies of the hydrogen bonding in
2-propanol~\cite{Schaal:JPCA104:265} and its hydrogen-bonded complexes~\cite{Evangelisti:PCCP19:568,
   Leon:PCCP16:16968}. Generally, the supersonic expansion provides beams of molecules at low
rotational temperatures. However, clusters can be formed~\cite{Ng:ACP:263, Hillenkamp:JCP118:8699,
   Scoles:MolBeam:1and2} due to the attractive forces between molecules. The temperature- and
pressure-dependence of its cluster formation was investigated~\cite{Leckenby:PRSL280:409,
   Fischer:JCP83:1471}. Other factors influencing the cluster formation in expanding supersonic jets
are the nozzle size and shape and the carrier gas~\cite{Scoles:MolBeam:1and2, Hagena:JCP5:1793,
   Jonkman:JPC89:4240}. Larger clusters are often fragile and then fragment upon excitation or
ionization before detection, rendering size-assignment from spectra ambiguous or even
impossible~\cite{Haberland:SS156:305, Schoellkopf:Science266:1345}.

For prospective studies of 2-propanol in chemical-reactions~\cite{Chang:Science342:98,
   Kilaj:NatComm9:2096} or diffractive-imaging experiments~\cite{Kuepper:PRL112:083002,
   Yang:Science361:64}, a cold and pure beam of 2-propanol separated from clusters as well as seed
gas is necessary~\cite{Filsinger:PCCP13:2076, Chang:IRPC34:557}. Such spatial separation of
molecular conformers was previously achieved using inhomogeneous electric
fields~\cite{Chang:IRPC34:557, Meerakker:CR112:4828}. Different species of complex molecules can be
spatially separated within a cold molecular beam by the electrostatic
deflector~\cite{Chang:IRPC34:557}, which was demonstrated in a number of pioneering experiments on
the separation of individual quantum states~\cite{Filsinger:JCP131:064309, Nielsen:PCCP13:18971,
   Horke:ACIE53:11965}, structural isomers~\cite{Filsinger:PRL100:133003, Kierspel:CPL591:130}, or
specific cluster sizes~\cite{Trippel:PRA86:033202, Teschmit:ACIE57:13775, You:JPCA122:1194,
   Johny:CPL721:149, Bieker:JPCA123:7486}. Here, the 2-propanol monomer is spatially separated from
the original molecular beam using the deflector and the purified samples are exploited in
femtosecond-laser ionization studies.

\section{Experimental Methods}
\begin{figure}[b]
   \includegraphics[width=\linewidth]{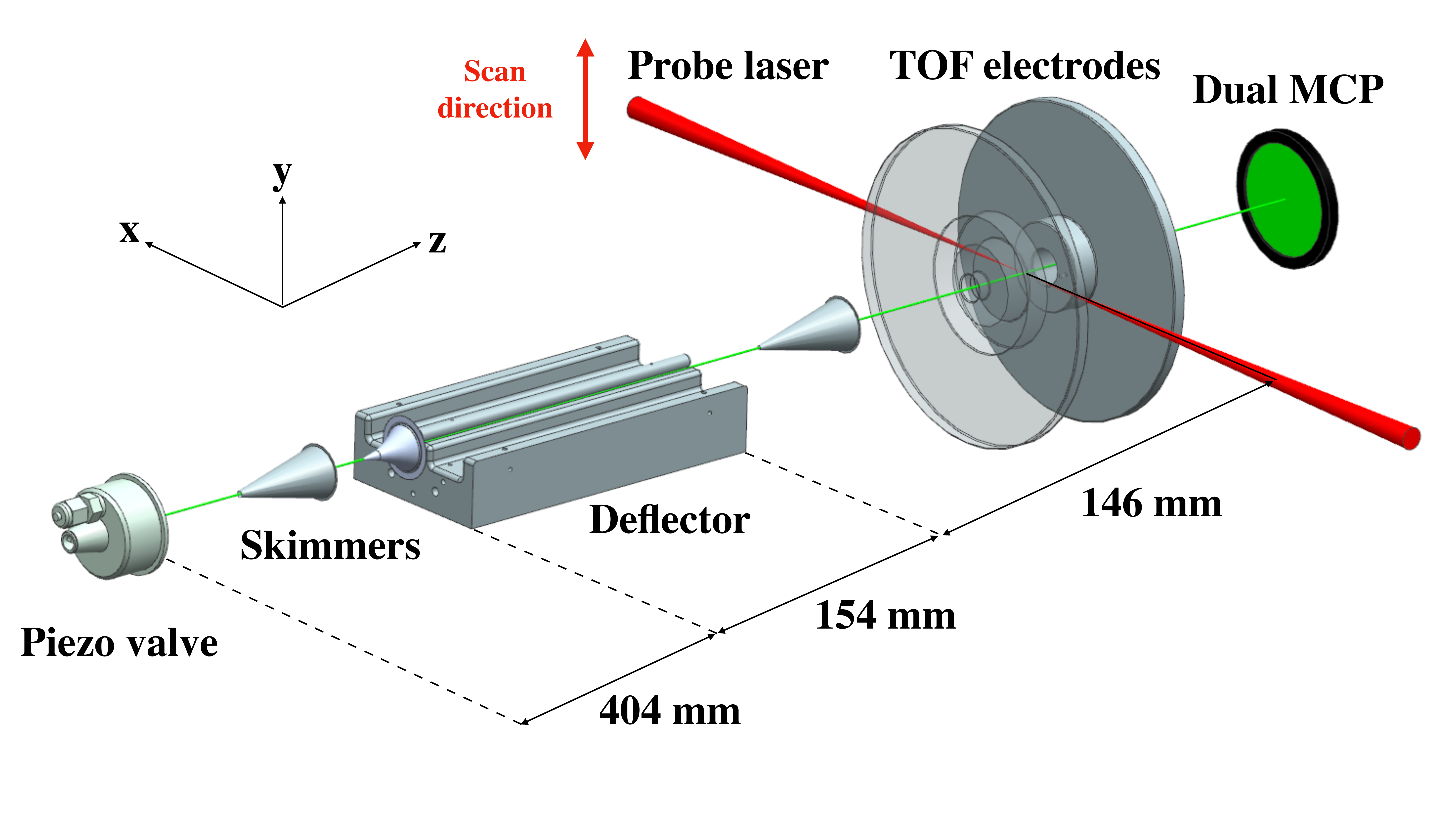}
   \caption{Schematic of the experimental setup, consisting of a pulsed valve, the electrostatic
      deflector, a femtosecond-pulse laser system, and a two-plate time-of-flight (TOF) mass
      spectrometer; see text for details.}
   \label{fig:expSetup}
\end{figure}
2-propanol was commercially obtained (Carl Roth GmbH, $>\!99~\%$) and used without further
purification. \autoref{fig:expSetup} shows a schematic of the experimental setup, similar to a
previously described one~\cite{Teschmit:ACIE57:13775}. Briefly, 2-propanol at room temperature is
seeded in 2~bar of helium and supersonically expanded into vacuum through a cantilever piezo
valve~\cite{Irimia:RSI80:113303} at a repetition rate of 20 Hz. The generated jet was differentially
pumped and collimated by two skimmers, which were placed 55~mm ($\varnothing=3$~mm) and 365~mm
($\varnothing=1.5$~mm) downstream of the valve. The resulting molecular beam was directed through
the electrostatic deflector, which was placed 404~mm downstream of the valve, before passing through
a third skimmer ($\varnothing=1.5$~mm). 2-propanol molecules were ionized by amplified femtosecond
laser pulses with a duration of 45~fs (full-width at half maximum, FWHM) with a spectrum centered
around 800~nm. By focusing the laser with pulse energies of 30~\uJ and 60~\uJ to 50~\um, we obtained
nominal peak intensities of $3\times10^{13}~\Wpcmcm$ and $7\times10^{13}~\Wpcmcm$, respectively. A
two-plate velocity map imaging (VMI)~\cite{Kienitz:CPC17:3740} was operated as a time-of-flight
(TOF) mass spectrometer (MS). Ions were counted through a single-shot detector readout and
centroiding algorithm~\cite{Trippel:MP111:1738}, and summed up to yield the mass spectra shown in
\autoref[a,~b]{fig:deflection}.

\section{Results and Discussion}
\label{sec:setup}
\begin{figure*}
   \includegraphics[width=\linewidth]{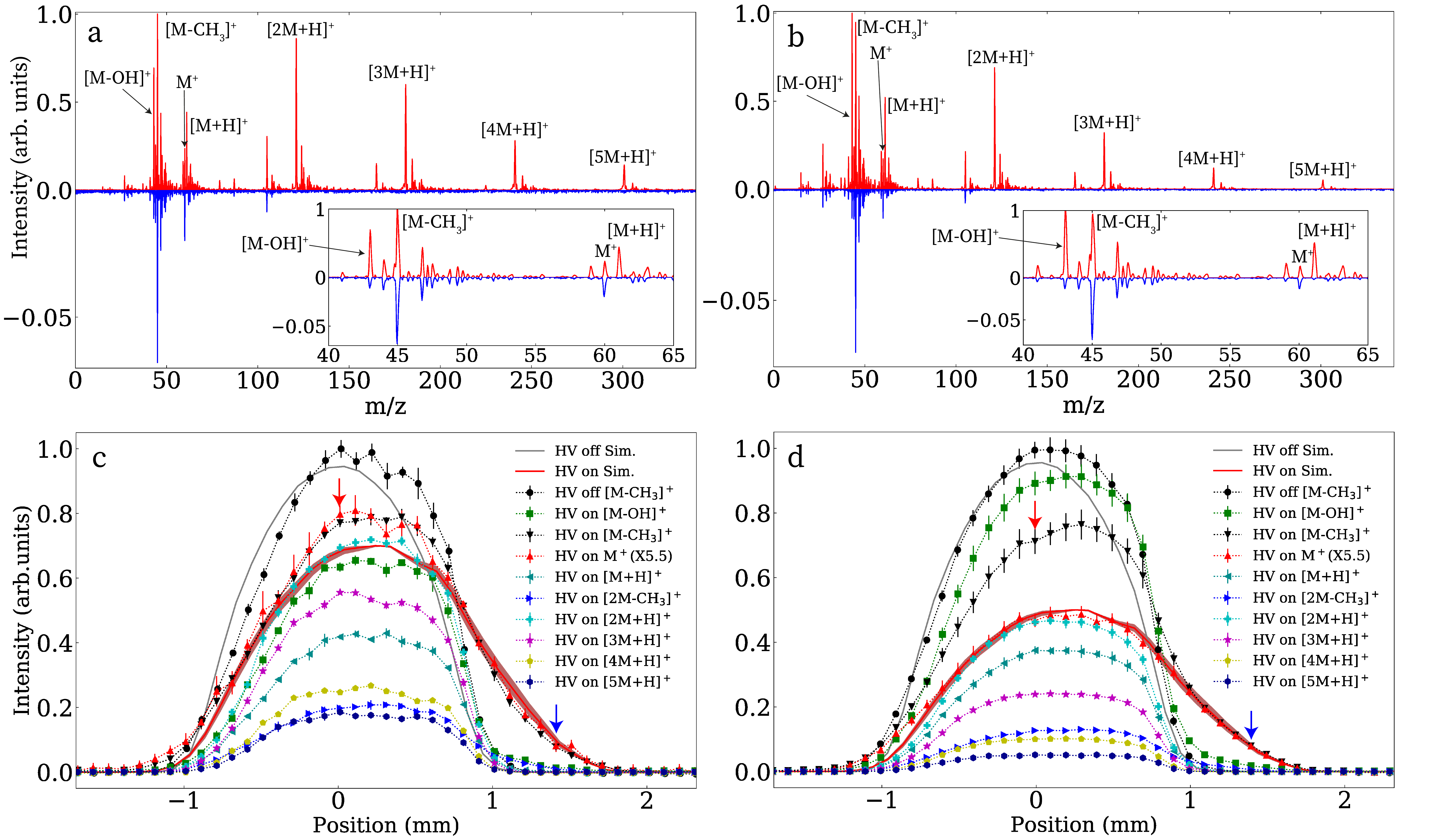}
   \caption{Mass spectra and vertical molecular-beam-density profiles with ionization-laser peak
      intensities of (a, c) $3\times10^{13}~\Wpcmcm$ and (b, d) $7\times10^{13}~\Wpcmcm$. (a) and
      (b) TOF-MS in the central part of the molecular beam with the deflector off (red) and at a
      position of $+1.4$~mm with a deflection voltage of 13~kV (blue). (c) and (d) 0 kV (circle) and
      13 kV vertical spatial-beam-profiles for both pulse energies. In the legends M refers to the
      monomer parent ion. Dashed lines show fragment cations of 2-propanol ([M-OH]$^{+}$ (green),
      [M-CH$_{3}$]$^{+}$ (black), [2M-CH$_{3}$]$^{+}$ (blue)), monomer cation M$^{+}$ (red),
      protonated 2-propanol [M+H]$^{+}$ (darkcyan) and its clusters up to $n=5$ ([2M+H]$^{+}$
      (cyan), [3M+H]$^{+}$ (magenta), [4M+H]$^{+}$ (yellow), [5M+H]$^{+}$ (darkblue)). Simulated
      deflection profiles of the direct 2-propanol monomer beam (gray) as well as the deflected
      2-propanol monomer beam (red) are shown as solid lines. The shaded (dark red) area depicts the
      error estimate of the 2-propanol simulation due to the temperature uncertainty. The red and
      blue arrows indicate the positions in the deflected beam where the mass spectra shown in (a)
      and (b) were measured.}
   \label{fig:deflection}
\end{figure*}
The mass spectra at peak intensities of $3\times10^{13}~\Wpcmcm$ and $7\times10^{13}~\Wpcmcm$ of the
direct (0 kV) and deflected (13 kV) molecular beams are shown in \autoref{fig:deflection}. The
spectrum of the direct beam shows 2-propanol fragment ions ([M-OH]$^{+}$ and [M-CH$_{3}$]$^{+}$),
monomer ions M$^{+}$, and protonated 2-propanol and its cluster ions [M$_n$+H]$^+$ up to $n=5$.
Here, M is the parent molecule and [M-OH]$^{+}$ and [M-CH$_{3}$]$^{+}$ specify fragments where M
lost OH or CH$_{3}$, respectively. The mass spectra in \autoref[a]{fig:deflection} and
\autoref[b]{fig:deflection} were normalized to their largest peak. Note that the mass spectra of the
direct and deflected beams are shown on different vertical scales for improved visibility. Larger
clusters were not detected due to the recorded TOF interval in the experiments. Note that the
clusters of 2-propanol in the interaction region are neutral and that the protonated-cluster ions in
the mass spectrum resulted from the ionization detection process~\cite{Johny:CPL721:149,
   Bieker:JPCA123:7486}.

The spatial vertical molecular-beam-density profiles for 2-propanol monomer ions, fragment ions,
protonated 2-propanol, and its cluster ions [M$_n$+H]$^+$ up to $n=5$, ionized with the laser peak
intensities of $3\times10^{13}~\Wpcmcm$ and $7\times10^{13}~\Wpcmcm$, are shown in
\autoref[c]{fig:deflection} and \autoref[d]{fig:deflection}. The direct and deflected profiles were
normalized to their largest signals. For better visibility, the M$^{+}$ deflection profile has been
scaled up by a factor of 5.5. When a voltage of 13~kV is applied to the deflector, the profiles of
fragment ions ([M-OH]$^{+}$ and [M-CH$_{3}$]$^{+}$), and the parent ion M$^{+}$ are shifted by
$+1.8$~mm at both peak intensities; the [2M-CH$_{3}$]$^{+}$ profile also has a tail with significant
deflection, which we ascribe to a fairly polar cluster, but this species deflects less and has only
a small population in the beam. The protonated 2-propanol and its cluster ions in
\autoref[c]{fig:deflection} and \autoref[d]{fig:deflection} do not deflect in the region of 1.1 to
1.8~mm, which shows that dimers and larger clusters generally deflect much less than the monomer and
the monomer ion signal originated from the monomer. In \autoref[a]{fig:deflection} and
\autoref[b]{fig:deflection}, the intensity of the cluster peaks is reduced at higher intensity. At
the same time, the relative intensity of the [M-OH]$^{+}$ peak is strongly increased, which suggests
that [M-OH]$^{+}$ is a major fragmentation product of these larger clusters.

2-propanol is a nearly symmetric oblate rotor, the rotational constants for \emph{trans} and
\emph{gauche} were obtained by microwave spectroscopy~\cite{Hirota:JPC83:1457}. The dipole-moment
components are $\mu_{a}=0,\mu_{b}=1.40~\text{D},\mu_{c}=0.73$~D for the
\emph{trans}~\cite{Kondo:JMS34:97} and
$\mu_{a}=1.114~\text{D},\mu_{b}=0.737~\text{D},\mu_{c}=0.813$~D for the \emph{gauche} conformer,
respectively~\cite{Hirota:JPC83:1457}. Their energy difference is smaller than 1~kJ/mol and the
isomerization barrier corresponding to rotation of the hydroxyl moiety is
low~\cite{Leon:PCCP16:16968}. The gauche conformer is more stable than the trans conformer and there
is a strong conformational relaxation of 2-propanol monomer from trans to
gauche~\cite{Ruoff:JCP93:3142}. Using the specified dipole-moment components and the known
rotational constants~\cite{Hirota:JPC83:1457}, the Stark energies and effective dipole moments of
both forms were calculated with our \textsc{CMIstark} software package~\cite{Chang:CPC185:339}. The
Stark-effect differences between the two conformers are too small for their separation in this
experiment. Furthermore, it is difficult to determine the ratio between the trans and gauche forms
at low temperatures as they are essentially identical. For the analysis in this work, we used only
the trans-conformer deflection simulations.

The simulated vertical molecular-beam profiles of the 2-propanol monomer are shown in
\autoref[c]{fig:deflection} and \autoref[d]{fig:deflection}. The Stark energies for all rotational
states up to $J=14$ were calculated using a basis of field-free rotational states up to
$J=30$~\cite{Chang:CPC185:339}. For every quantum state, $1\times10^{5}$ trajectories were
calculated. The initial beam temperature that described the experimental observations best was
determined to be $3.5(5)$~K. The shaded (dark red) area depicts the error estimate of the 2-propanol
simulation due to the temperature uncertainty.

In \autoref[c]{fig:deflection} the deflected profiles (dashed lines) of monomer (red triangle up)
and [M-CH$_{3}$]$^{+}$ ions (black triangle down) matched very well over the whole deflection
region, indicating that [M-CH$_{3}$]$^{+}$ and M$^{+}$ both originated from the parent molecule.
However, at higher peak ionization-laser intensity, \autoref[d]{fig:deflection}, the profiles of
monomer and [M-CH$_{3}$]$^{+}$ ions matched only in the region of 1.1 to 1.8~mm. The higher
[M-CH$_{3}$]$^{+}$ signal in the [-1,1] regions indicates the contribution of larger clusters to
this fragment in the ion signals. Nevertheless, this behavior nicely confirms the selection of
monomers from the expansion in the deflected part at positions larger than 1.1~mm.

\begin{figure}
   \includegraphics[width=\linewidth]{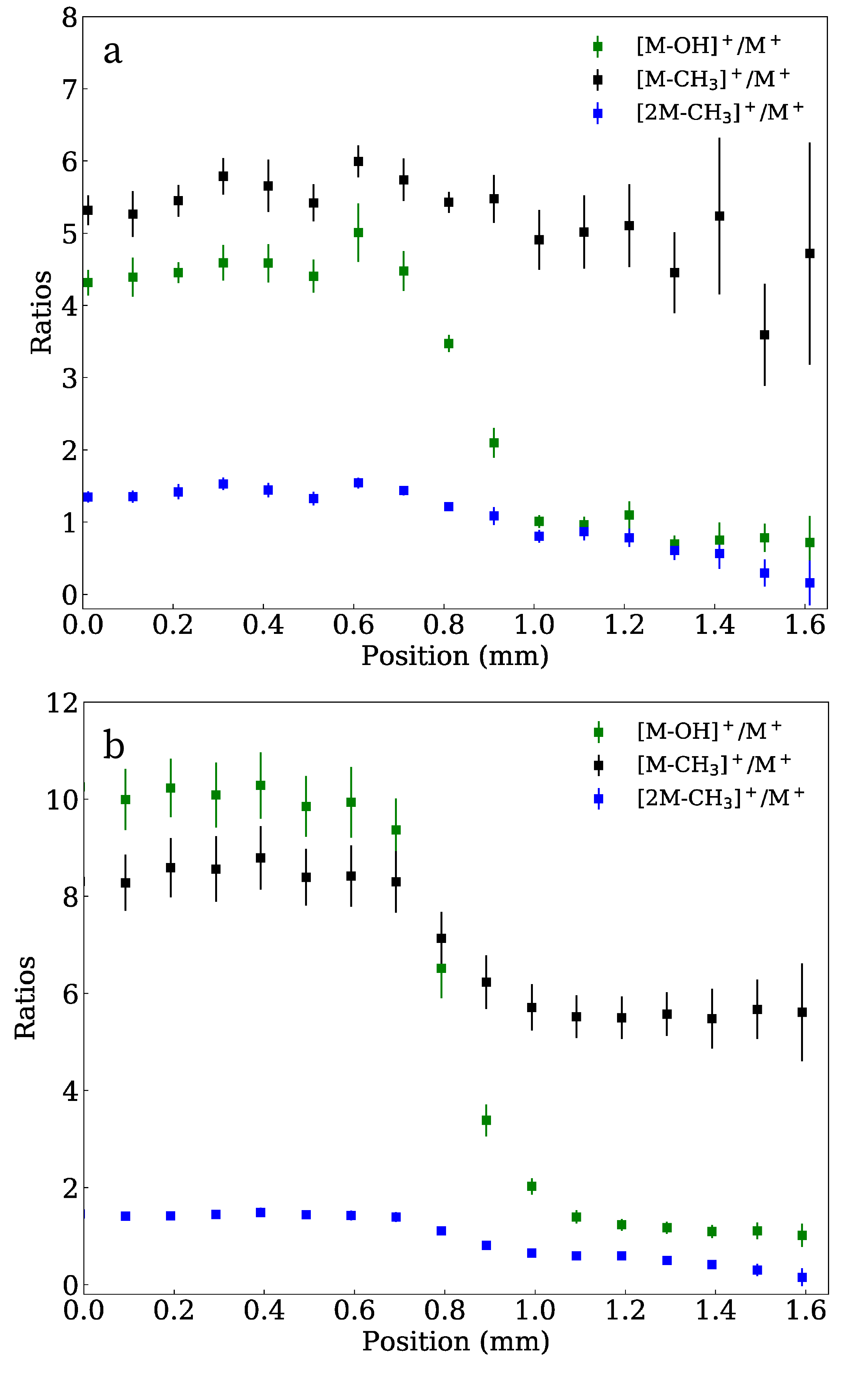}
   \caption{The ratios of different fragment ions [M-OH]$^{+}$ (green), [M-CH$_{3}$]$^{+}$ (black),
      [2M-CH$_{3}$]$^{+}$ (blue) and M$^{+}$ ion in the deflected region with peak intensities of
      (a) $3\times10^{13}~\Wpcmcm$ and (b) $7\times10^{13}~\Wpcmcm$.}
   \label{fig:ratio}
\end{figure}
The deflected-beam mass spectra in \autoref[a]{fig:deflection} and \autoref[b]{fig:deflection}
mainly contained peaks corresponding to [M-OH]$^{+}$, [M-CH$_{3}$]$^{+}$, M$^{+}$ and
[2M-CH$_{3}$]$^{+}$ at both laser peak intensities. The fragments were caused by the fs-pulse
ionization process, however, no hydrogen ions were observed in the TOF spectrum in
\autoref[a]{fig:deflection} due to the low laser peak intensity. The ratios of different fragments
and M$^{+}$ ions from the center position to 1.6~mm are shown in \autoref{fig:ratio}. For both laser
peak intensities, the ratios for [M-OH]$^{+}$, and [M-CH$_{3}$]$^{+}$ ions in the deflected region
from 1.1~mm to 1.6~mm are constant, which suggests [M-OH]$^{+}$, [M-CH$_{3}$]$^{+}$ and M$^+$ ions
come from the same parent, the monomer. However, the ratio for [2M-CH$_{3}$]$^{+}$ decreases as the
position increases in \autoref[a]{fig:ratio} and \autoref[b]{fig:ratio}, indicating that this peak
originates from the dimer or larger clusters, which are deflected out of the original beam, but less
than the monomer.

Following the strong-field/multi-photon ionization process of 2-propanol, C--C and C-O bonds of the
monomer were broken. This suggests that the monomer has two fragmentation channels, M$\rightarrow$
[M-CH$_{3}$]$^{+}$ and M$\rightarrow$ [M-OH]$^{+}$.

Based on the intensity of the fragments of the 2-propanol in the deflected beam,
\autoref[a]{fig:deflection} and \autoref[b]{fig:deflection}, the purity of the intact 2-propanol
monomer was derived as the ratio of the sum of the signals due to M$^{+}$ and its known fragments
[M-OH]$^{+}$ and [M-CH$_{3}$]$^{+}$ to the sum of all signals in the mass spectrum. The 2-propanol
monomer fraction was $4(1)~\%$ in the center of the direct beam and $90(4)~\%$ at $+1.4$~mm in the
deflected beam. This is a nearly 22-fold increase in the fractional density of 2-propanol from the
direct beam to the deflected beam.

By comparing the intensities of [M-CH$_{3}$]$^{+}$ and [M-OH]$^{+}$ in \autoref[a]{fig:ratio} and
\autoref[b]{fig:ratio}, the intensity ratio of the M$\rightarrow$ [M-CH$_{3}$]$^{+}$ channel to the
M$\rightarrow$ [M-OH]$^{+}$ channel was obtained as $\ordsim5$ for both peak intensities in the
region from 1.1 mm to 1.6~mm. The fragmentation ratios of the monomer, defined as the intensities of
the fragments divided by the sum of intensities of the monomer and its related fragments, were
estimated as $83~\%$ and $89~\%$ at the lower and higher peak intensity, respectively.

The beam density was estimated based on the analog detector current signal calibrated to a
single-ion hit. Approximately 7~ions/shot were created in the deflected beam ($+1.4$~mm) at a laser
peak intensities of $7\times10^{13}~\Wpcmcm$. Assuming a typical detection efficiency of 0.5 for the
MCP detector, a molecular-beam width of 1~mm, and a strong-field-ionization probability of 1, a beam
density of $\ordsim7\times10^{6}$~cm$^{-3}$ was obtained for the 2-propanol monomer.

\section{Conclusions}
A high-purity beam of 2-propanol monomer was produced through the spatial separation of the monomer
from its clusters and the seed gas using the electrostatic deflector. The purity and beam density of
2-propanol monomer were $90(4)\%$ and $7\times10^{6}~\text{cm}^{-3}$ in this deflected part of the
molecular beam. The 45~fs laser-pulse ionization of 2-propanol with peak intensities of
$3\times10^{13}~\Wpcmcm$ and $7\times10^{13}~\Wpcmcm$ was studied. The 2-propanol monomer showed two
fragmentation channels in the strong-field ionization process and the ratio of M$\rightarrow$
[M-CH$_{3}$]$^{+}$ to $\text{M}\rightarrow\text{[M-OH]}^{+}$ was estimated to be $\ordsim5$. The
fragmenting fractions of the monomer were estimated to be $83~\%$ and $89~\%$ at the lower and
higher peak intensities, respectively. The produced intense, cold, and pure 2-propanol monomer beam
is well-suited for further investigations, such as diffractive imaging~\cite{Kuepper:PRL112:083002,
   Barty:ARPC64:415}, chemical reaction~\cite{Chang:Science342:98} and
combustion~\cite{Esarte:Energy43:37} studies.

\section*{Acknowledgments}
\label{sec:Acknowledgments}
This work has been supported by the Clusters of Excellence ``Center for Ultrafast Imaging'' (CUI,
EXC~1074, ID~194651731) and ``Advanced Imaging of Matter'' (AIM, EXC~2056, ID~390715994) of the
Deutsche Forschungsgemeinschaft (DFG) and by the European Research Council under the European
Union's Seventh Framework Program (FP7/2007-2013) through the Consolidator Grant COMOTION
(ERC-Küpper-614507). J.W.\ and L.H.\ acknowledge fellowships within the framework of the
Helmholtz-OCPC postdoctoral exchange program and J.O.\ gratefully acknowledges a fellowship by the
Alexander von Humboldt Foundation.

\bibliography{iso-propanol-deflection}%
\onecolumngrid
\end{document}